\documentclass[11pt]{article}
\usepackage{moriond,epsfig}

\def\be{\begin{equation}}
\def\ee{\end{equation}}
\def\bea{\begin{eqnarray}}
\def\eea{\end{eqnarray}}

\newcommand{\raw}{\rightarrow}

\newcommand{\evb}{ {\rm eV$^2$} } 
 
\newcommand{\gev}{ {\rm GeV} }

\newcommand{\PCPV}{
\begin{picture}(22,10)
\put(8,-2){\line(2,1){12}}
\put(0,0){$P_{CP}$}
\end{picture}}
\newcommand{\PCPC}{
\begin{picture}(22,10)
\put(0,0){$P_{CP}$}
\end{picture}}
\begin{document}
\vspace*{4cm}
\title{Four Neutrino Mass Spectra at the Neutrino Factory\footnote{Presented by A. Donini}}

\author{A. Donini$^{(1,2)}$~and~D. Meloni$^{(2,1)}$}

\address{
$^{(1)}$ Istituto Nazionale di Fisica Nucleare, 
Sezione di Roma, I-00185 Rome, Italy \\
$^{(2)}$ Universit\`a di Roma ``La Sapienza'',
I-00185 Rome, Italy \\}

\maketitle\abstracts{
We study the physical reach of a Neutrino Factory in the 2+2 and 3+1
Four Neutrino mixing scenarios, with similar
results for the sensitivity to the mixing angles. Huge CP-violating 
effects can be observed in both schemes with a near, $O(10)$ Km, detector of 
$O(10)$ Kton size in the $\nu_\mu \to \nu_\tau$ channel. 
A smaller detector of 1 Kton size can still observe very large effects
in this channel. 
}

\section{Introduction}

Indications in favour of neutrino oscillations have been obtained 
both in solar neutrino \cite{solar} and atmospheric neutrino \cite{atmospheric} 
experiments with $\Delta m^2_{sol} \leq 10^{-4}$ \evb 
and $\Delta m^2_{atm} \sim 10^{-3}$ \evb.
The LSND data \cite{Aguilar:2001ty} would indicate 
a $\bar \nu_\mu \to \bar \nu_e$ oscillation with a third 
neutrino mass difference: $\Delta m_{LSND}^2 \sim 0.3 - 6$ \evb. 
If MiniBooNE \cite{Church:1997jc} confirms the LSND results 
we would therefore face three independent evidence for neutrino oscillations 
characterized by squared mass differences quite well separated. 
To explain the whole ensemble of data at least four different light 
neutrino species are needed. 

There are two classes of four neutrino spectra: 
three almost degenerate neutrinos and an isolated fourth one (the {\it 3+1 scheme}), 
or two pairs of almost degenerate neutrinos divided by the large
LSND mass gap (the {2+2 scheme}). Although the latter is still favoured \cite{Grimus:2001mn},
the new analysis of the experimental data \cite{Aguilar:2001ty} results 
in a shift of the allowed region towards smaller values of the mixing angle, 
$\sin^2 (2 \theta)_{LSND}$, reconciling the 3+1 scheme with exclusion 
bounds \cite{exclusionLSND,3+1citations}. 

Four neutrino oscillations imply a Maki-Nakagawa-Sakata (MNS) $4 \times 4$
mixing matrix, with six rotation angles $\theta_{ij}$ and three phases $\delta_i$
(for Dirac-type neutrinos).
This large parameter space is actually reduced to a smaller subspace whenever 
some of the mass differences become negligible. Consider the measured hierarchy in the 
mass differences, $\Delta m^2_{sol} \ll \Delta m^2_{atm} \ll \Delta m^2_{LSND}$
and define $\Delta_{ij} = \Delta m^2_{ij} L / (4 E_\nu)$.
At short distance, $L = O(1)$ Km, for neutrinos up to $O (10)\, \gev$, 
$ ( \Delta_{sol}, \Delta_{atm}) \ll 1$ and $\Delta_{LSND} = O(1)$ and we can
safely neglect the solar and atmospheric mass differences. 
In the 2+2 scheme the rotation angles in the $(1-2)$ and $(3-4)$ planes
become irrelevant in oscillation experiments, together with
two CP-violating phases, and the parameter space reduces to
4 rotation angles and 1 phase only. In the 3+1 scheme the rotations 
in the whole three-dimensional subspace $(1-2-3)$
are irrelevant for oscillation experiments, and the physical parameter
space contains three rotation angles and no phases. 
When considering CP-violating phenomena at least two mass
differences should be taken into account.
In this approximation, regardless of the scheme, 
the parameter space contains 5 angles and 2 phases. 

In the 2+2 scheme, the following parametrization was adopted \cite{Donini:1999jc}:
\be
U_{MNS} = U_{14} (\theta_{14}) \; 
            U_{13} (\theta_{13}) \; 
            U_{24} (\theta_{24}) \; 
            U_{23} (\theta_{23}\, , \, \delta_3) \; 
\times
            U_{34} (\theta_{34}\, , \, \delta_2) \; 
            U_{12} (\theta_{12}\, , \, \delta_1);
\label{2+2param}
\ee
in the 3+1 scheme, the following parametrization \cite{Donini:2001xy}
shares the same virtues of eq.~(\ref{2+2param}):
\bea
U_{MNS} = U_{14} (\theta_{14}) \; 
            U_{24} (\theta_{24}) \; 
            U_{34} (\theta_{34}) \; 
\times 
            U_{23} (\theta_{23}\, , \, \delta_3) \; 
            U_{13} (\theta_{13}\, , \, \delta_2) \; 
            U_{12} (\theta_{12}\, , \, \delta_1).
\label{3+1param}
\eea
In the one-mass dominance approximation, the unphysical angles and phases 
automatically decouple in both cases. 

A Neutrino Factory \cite{Geer:1998iz,DeRujula:1999hd} is perfectly suited to explore 
this large parameter space, hopefully including the discovery of 
leptonic CP violation \cite{machines}.
A comparison of the physical reach of a Neutrino Factory 
in the 2+2 and 3+1 schemes has been extensively presented elsewhere 
\cite{Donini:2001xy} and will be summarized here.
We shall consider in what follows as a ``reference set-up'' a neutrino beam 
resulting from the decay of $n_\mu = 2 \times 10^{20}$ unpolarized 
positive and/or negative muons per year. 
The collected muons have energy $E_\mu$ in the range $10 - 50$ GeV. 

As the dominant signals are expected to peak at $L/E_\nu \sim 1/\Delta m^2_{LSND}$, 
most of the CP-conserving parameter space can be explored in short baseline experiments 
($L \sim 1$ Km) with a small size detector with $\tau$ tracking and ($\mu, \tau$) 
charge identification capability. We consider an hypothetical 1 ton detector
with constant background $B$ at the level of $10^{-5}$ of the expected number of
charged current events and a constant reconstruction efficiency 
$\epsilon_\mu = 0.5$ for $\mu^\pm$ and $\epsilon_\tau = 0.35$ for $\tau^\pm$ 
(neutrinos with $E_\nu \leq 5$\gev have not been included).
To extend our analysis to the CP-violating parameter space we consider 
an hypothetical $10$ Kton detector, located a bit farther 
from the neutrino source, at $L = O(10-100)$ Km. 

%%%%%%%%%%%%%%%%%%%%%%%%%%%%%%%%%%%%%%%%%%%%%%%%%%%%%%%%%%%%%%%%

\section{Sensitivity reach of the Neutrino Factory}
\label{sensi}

We follow a conservative (or even ``pessimistic'') hypothesis \cite{Donini:1999jc} 
and consider the four gap-crossing angles in 
the 2+2 scheme, $\theta_{13}, \theta_{14},\theta_{23}$ and $\theta_{24}$, 
to be equally small (i.e. less than $10^\circ$),
with the possible exception of one angle free to vary in some interval. 
The remaining angles $\theta_{12}$ and $\theta_{34}$ are the
solar and atmospheric mixing angles in the two-family parametrization, respectively. 
In the 3+1 scheme we restrict to one of the allowed regions 
\cite{3+1citations}, $\Delta m^2_{34} = 0.9 $ \evb, $\sin^2 (2 \theta)_{LSND} 
\simeq 2 \times 10^{-3}$, for simplicity, and we take equally small
gap-crossing angles $\theta_{i4}$. The remaining angles, 
$\theta_{12}, \theta_{23}$ and $\theta_{13}$ can be obtained by the combined 
analysis of solar and atmospheric data in the three-family parametrization \cite{Fogli:1999yz}.

Our results \cite{Donini:2001xy} show that the considered set-up can severely 
constrain the whole four-family model 
CP-conserving parameter space, both in the 2+2 scheme and 3+1 scheme.
In the former, the sensitivity reach to all gap-crossing angles in the 
LSND-allowed region is at the level of $\sin^2 \theta \geq 10^{-6} - 10^{-4}$, 
depending on the specific angle considered. In the latter 
the sensitivity reach is at the level of $\sin^2 \theta \geq 10^{-5} - 10^{-3}$, 
slightly less than in the 2+2 case.

This results can be easily understood in terms of a simple power counting argument 
\cite{Donini:2001xy}. In the third column of Tab. \ref{tab:power} 
we report the leading order in $\epsilon$ for the CP-conserving oscillation
probabilities $\PCPC$ in the three-family model and in both schemes
of the four-family model. In three families, the small parameter is $s_{13} \sim \epsilon$.
In four families we consider equally small LSND gap-crossing angles: 
$ s_{13} = s_{14} = s_{23} = s_{24} \sim \epsilon $ for the 2+2 scheme;
$ s_{14} = s_{24} = s_{34} = \epsilon $ for the 3+1 scheme. In this 
last case we take $s_{13} \sim \epsilon$, also. 
Notice that the appearance transition probabilities in the 2+2 scheme are generically 
of $O(\epsilon^2)$, with the only exception of $\nu_\mu \to \nu_\tau$. 
In the 3+1 scheme, on the contrary, they are all $O (\epsilon^4)$. 
This explains the (slight) decrease in the sensitivity in the 3+1 scheme
with respect to the 2+2 scheme. 

\begin{table}
\begin{center}
\begin{tabular}{||c||c|c|c|c||}
\hline
\hline
Scheme        & Transition & $\PCPC $ & $ \PCPV $ & $ A / \Delta A $ \\
\hline
\hline
              & $\nu_e \raw \nu_\mu $ & $\epsilon^2$ & $\epsilon$ & $ O(1) $  \\ 
Three-family  & $\nu_e \raw \nu_\tau$ & $\epsilon^2$ & $\epsilon$ & $ O(1) $  \\
            & $\nu_\mu \raw \nu_\tau$ &            1 & $\epsilon$ & $ O(\epsilon) $  \\
\hline
\hline
              & $\nu_e \raw \nu_\mu $ & $\epsilon^2$ & $\epsilon^2$ & $ O(\epsilon) $  \\ 
2+2         & $\nu_e \raw \nu_\tau$ & $\epsilon^2$ & $\epsilon^2$ & $ O(\epsilon) $  \\
            & $\nu_\mu \raw \nu_\tau$ & $\epsilon^4$ & $\epsilon^2$ & $ O(1) $  \\
\hline
\hline
              & $\nu_e \raw \nu_\mu $ & $\epsilon^4$ & $\epsilon^3$ & $ O(\epsilon) $  \\ 
3+1         & $\nu_e \raw \nu_\tau$ & $\epsilon^4$ & $\epsilon^3$ & $ O(\epsilon) $  \\
            & $\nu_\mu \raw \nu_\tau$ & $\epsilon^4$ & $\epsilon^2$ & $ O(1) $  \\
\hline
\hline
\end{tabular}
\caption{{ \it Small angles suppression in the CP-conserving 
and CP-violating oscillation probabilities, and in the signal-to-noise 
ratio of the CP asymmetries, in the three-family model and in both 
four-family model mass schemes.}}
\label{tab:power}
\end{center}
\end{table}

\section{CP-violating Observables}
\label{cp-viola}

In the four-family model we can consider CP-violating observables whose overall size 
does not depend on $\Delta_{sol}$ (as in three families) but on $\Delta_{atm}$. 
Large CP-violating effects are therefore possible in this case.
Finally, the CP-violating observables are maximized for 
$L/E_\nu \sim 1 / \Delta m^2_{LSND} = O(10)$ Km (for neutrinos of $E_\nu = O(10)$ GeV) 
and matter effects are therefore completely negligible.

We consider the neutrino-energy integrated quantity \cite{DeRujula:1999hd}: 
\begin{equation}
{\bar A}^{CP}_{ \alpha \beta} (\delta) = 
\frac{ \{ {N[l_\tau^-]}/{N_o[l_\mu^-]} \}_{+} 
       - \{N[l_\tau^+]/N_o[l_\mu^+]    \}_{-} }{ 
       \{  N[l_\tau^-]/N_o[l_\mu^-]    \}_{+} 
       + \{N[l_\tau^+]/N_o[l_\mu^+]     \}_{-}} \; ,
\label{intasy}
\end{equation}
where $N[l_\tau^\pm]$ is the number of taus due to oscillated neutrinos and 
$N_o[l_\mu^\pm]$ is the expected number of muons in the absence of oscillations.
In order to quantify the significance of the signal, we compare the 
value of the integrated asymmetry with its error, $\Delta {\bar A}^{CP}_{ \alpha \beta}$, 
in which we include the statistical error and a conservative background estimate at
the level of $10^{-5}$, and subtract the matter induced asymmetry
${\bar A}^{CP}_{ \alpha \beta}(0^\circ)$.

\begin{figure}[h!]
\begin{center}
\begin{tabular}{cc}
\epsfxsize6.5cm\epsffile{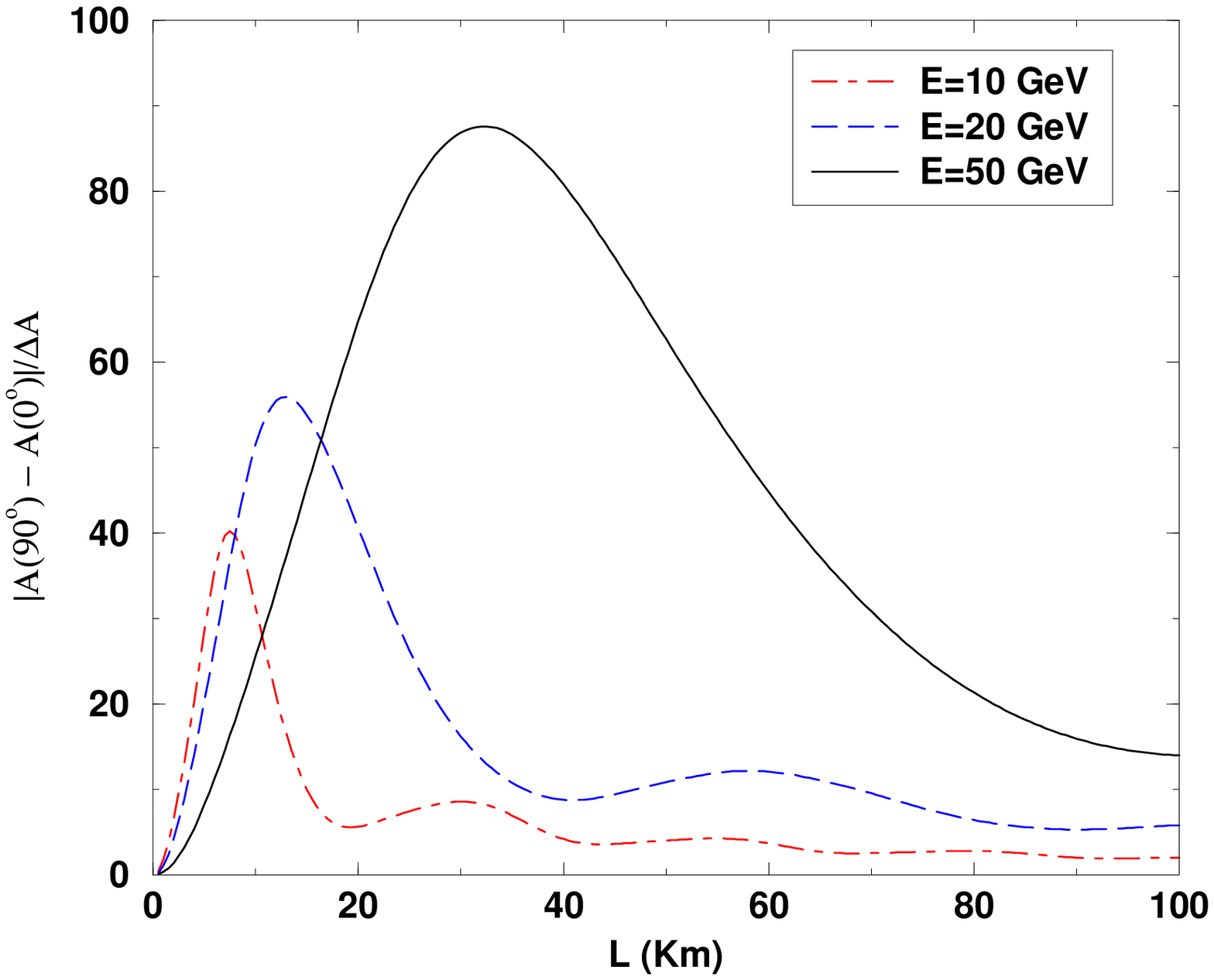} & 
\epsfxsize6.5cm\epsffile{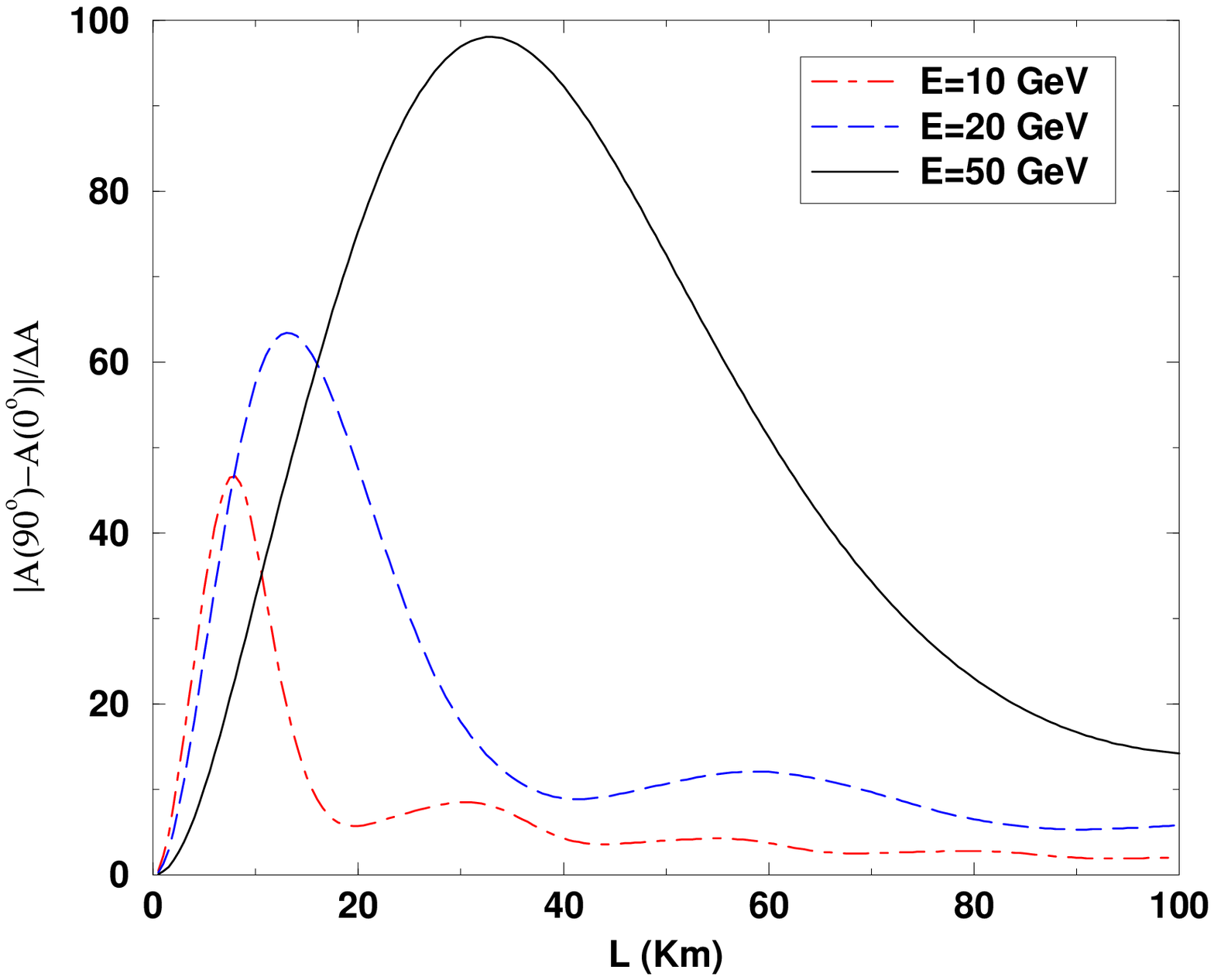}  \\
\end{tabular}
\caption{{\it Signal over statistical uncertainty for maximal
CP violation in the $\nu_\mu \to \nu_\tau$ channel 
in the 2+2 scheme (left) and in the 3+1 scheme (right), 
as a function of the baseline $L$, for three values of the
parent muon energy, $E_\mu = 10, 20$ and $50$ GeV. }}
\label{fig:mutaucp}
\end{center}
\end{figure}

In Fig. \ref{fig:mutaucp} we show the signal-to-noise ratio 
of the subtracted integrated CP asymmetry in the $\nu_\mu \to \nu_\tau$ channel
for the 2+2 (left) and the 3+1 (right) schemes, respectively. 
In both cases, for $E_\mu = 50$ \gev, $\sim 100$ standard deviations are attainable 
at $L \simeq 30 - 40$ Km. For a detector of size $M$, a reduction factor
$\propto 1/ \sqrt{M}$ should be applied. Therefore, for an OPERA-like $O(1)$ Kton detector 
we still expect large CP-violating effects in the $\nu_\mu \to \nu_\tau$ channel.
The other two channels, $\nu_e \to \nu_\mu, \nu_\tau$ give a much smaller significance 
in both schemes. 

The real gain with respect to the three-family model \cite{goldenetal}
is that the small solar mass difference, that modules the overall size
of the CP-violating asymmetry, is traded with the much larger atmospheric mass
difference. The optimal channel to observe CP violation 
is the $\nu_\mu \to \nu_\tau$ channel. This result can be easily understood looking at 
Tab. \ref{tab:power}. In the fourth and fifth columns we report the leading order 
in $\epsilon$ for the different CP-violating oscillation probabilities $\PCPV$
and for the related signal-to-noise ratio of the CP asymmetries (remind that $A / \Delta A$ 
is proportional to $\PCPV / \sqrt{\PCPC}$). Notice that in the three-family model 
the $\nu_e \to \nu_\mu, \nu_\tau$ channels have a signal-to-noise ratio of the 
corresponding CP asymmetry of $O(1)$ in the small angles. On the contrary, in both the 2+2 
and 3+1 four-family model, it is the $\nu_\mu \to \nu_\tau$ channel to be of $O(1)$ 
in the small angles, thus justifying {\em a posteriori} our results.

\section*{Acknowledgments}
A.~D. thanks the Organizer Committee 
for the always stimulating atmosphere of the Moriond conference.

\end{document}